\def\lsim{\raise0.3ex\hbox{$<$\kern-0.75em\raise-1.1ex\hbox{$\sim$}}} 
\def\gsim{\raise0.3ex\hbox{$>$\kern-0.75em\raise-1.1ex\hbox{$\sim$}}}
\documentclass[12pt]{JHEP3}
\usepackage{graphicx}

\title{The~QCD~transition~temperature:~results~with physical masses in the continuum limit II.}

\author{Yasumichi Aoki\\
RIKEN BNL Research Center, Brookhaven National Laboratory,
Upton, NY 11973, USA
}
\author{Szabolcs Bors\'anyi\\
Bergische Universit\"at Wuppertal, D-42119 Wuppertal, Germany\\
}
\author{Stephan D\"urr\\
Forschungszentrum J\"ulich, D-52425 J\"ulich, Germany\\
NIC/DESY Zeuthen Forschungsgruppe, D-15738 Zeuthen, Germany
}
\author{Zoltan Fodor\\
NIC/DESY Zeuthen Forschungsgruppe, D-15738 Zeuthen, Germany\\
Bergische Universit\"at Wuppertal, D-42119 Wuppertal, Germany\\
Forschungszentrum J\"ulich, D-52425 J\"ulich, Germany\\
Institute for Theoretical Physics, E\"otv\"os University, H-1117 Budapest, Hungary
}
\author{Sandor D. Katz\\
Bergische Universit\"at Wuppertal, D-42119 Wuppertal, Germany\\
Institute for Theoretical Physics, E\"otv\"os University, H-1117 Budapest, Hungary
}
\author{
Stefan Krieg\\
Bergische Universit\"at Wuppertal, D-42119 Wuppertal, Germany\\
Center for Theoretical Physics, Massachusetts Institute of Technology,
Cambridge, Massachusetts 02139, USA
}
\author{
Kalman Szabo \email{szaboka@general.elte.hu}\\
Bergische Universit\"at Wuppertal, D-42119 Wuppertal, Germany
}

\abstract{ We extend our previous study [Phys. Lett. B643 (2006) 46] of the
cross-over temperatures ($T_c$) of QCD.  We improve our zero temperature
analysis by using physical quark masses and finer lattices. In addition to the
kaon decay constant used for scale setting we determine four quantities (masses
of the $\Omega$ baryon, $K^*(892)$ and $\phi(1020)$ mesons and the pion 
decay constant)
which are found to agree with experiment. This implies that --independently 
of which of these quantities is used to set the overall scale-- the
same results are obtained within a few percent. At finite temperature we use
finer lattices down to $a\lsim 0.1$~fm ($N_t=12$ and $N_t=16$ at one point).
Our new results confirm completely our previous findings. We compare the
results with those of the 'hotQCD' collaboration.}


\preprint{WUB-09-01\\ ITP-Budapest 644\\RBRC-782}

\begin{document}
\section{Introduction}

There is a continuously high interest in determining properties of the high
temperature quark gluon matter. One of the major goals is to determine the
temperature scale, where the ordinary, hadronic matter is supposed to undergo a
transition to the high temperature phase. Since this transition seems to be a
continuous one \cite{Aoki:2006we}, there is no unambiguous temperature,
where the transition takes place. In general different observables may have
their characteristic points (e.g. peak position, inflection point) at different
temperature values. These temperatures are completely well defined and in
principle can be calculated with an arbitrary precision.

Current lattice simulations tend to disagree on these characteristic
temperature scales. On the one hand 
the published results of 
the RBC-Bielefeld collaboration
found \cite{Cheng:2006qk}
\begin{eqnarray}
\label{eq:tchot}
T_c=192(4)(7) {\rm ~MeV}
\end{eqnarray}
for the transition temperature. By considering different observables they
obtained transition temperature values that were consistent with each other.
In later works of the group (which has been enlargened to 'hotQCD' collaboration
in the meantime) the analysis has been extended to other fermion actions and
smaller lattice spacings \cite{Detar:2007as,Karsch:2007dt,Karsch:2008fe}. The
results presented in these works seem to confirm those of \cite{Cheng:2006qk}, in particular
\cite{Karsch:2007dt} concluded as: "The preliminary results of the hotQCD collaboration
indicate that the crossover region for both deconfinement and chiral symmetry
restoration lie in the range $T=(185$-$195$)~MeV". 

On the other hand the results that we presented in \cite{Aoki:2006br} are quite
different. Different observables led to significantly different transition
temperatures and these temperature values were considerably lower than the
values of the 'hotQCD' collaboration. For example for the transition temperature
defined by the peak position of the 
renormalized 
chiral susceptibility we obtained
\begin{eqnarray}
T_c(\chi_{\bar{\psi}\psi})=151(3)(3) {\rm ~MeV},
\end{eqnarray}
which is more than 20\% lower than the transition temperature of
\cite{Cheng:2006qk} (see Equation \ref{eq:tchot}).
The differences between the findings of the collaborations can be made even
more transparent and thus more disturbing by comparing the temperature
dependence of the observables. We have found discrepancy in all quantities that
we have considered so far, so it will be most probably present in the equation of
state, too.

Relating the above temperature scales to experimental observables of heavy-ion
collisions is a highly nontrivial task. Among other things one has to take into
account that most lattice calculations are carried out with periodic boundary
condition, which is convenient for the computations, but rather far from the
experimental setup.  An exploratory quenched 
study suggests \cite{Bazavov:2007zz} that
critical temperatures with realistic boundary conditions can be up to 30~MeV
larger than the values, which are measured in conventional lattice
calculations.

The aim of the present paper is to improve our previous results
\cite{Aoki:2006br} and to find some hints for the origin of the
discrepancies
discussed above. 
We present here three significant improvements:
\begin{itemize}

\item we extend our zero temperature simulations by simulating directly with
the physical values of the quark masses,

\item in order to verify that our results are independent of the physical
quantity we choose to set the scale we measured five experimentally
well-known quantities,

\item we extend our finite temperature simulations by taking an even smaller
lattice spacing ($N_t=12$ and at one point even $N_t$=16) than the smallest 
one we had in \cite{Aoki:2006br}.

\end{itemize}
The zero temperature results are presented in Section \ref{se:zerot}. The
finite temperature results are to be found in Section \ref{se:fint}, where a
comparison with the latest results of the 'hotQCD' collaboration is also done.

\section{Zero temperature simulations}
\label{se:zerot}

The primary role of zero temperature simulations is that they are used to
convert the dimensionless temperature of the lattice to physical units.
Therefore, when looking for systematic errors, one has to pay as much attention
to these simulations as to the finite temperature ones. In addition, zero
temperature runs are used to renormalize certain quantities in order to obtain
a meaningful continuum limit. Using these zero temperature simulations one can
also obtain the so called Lines of Constant Physics (LCP), which are
constraints among the lattice parameters. In our case the LCP tells us how to
tune the bare light quark masses ($m_{ud}$) and the bare strange quark mass
($m_s$) as the function of the gauge coupling ($\beta$) so that certain
hadronic quantities on the lattice take the same values as in the experiments.
In \cite{Aoki:2006br} we have determined the LCP using three hadronic
quantities: the pion and kaon masses and the kaon decay constant. When we say
that the
light or strange quark masses are set to their physical values, we mean that
they are on this LCP ($m_{ud}^{\rm LCP}$ or $m_{s}^{\rm LCP}$).

One shortcoming of 
essentially all lattice calculations these days
is that the zero temperature
runs were done at nonphysical light quark masses, only the strange quark mass
was fixed to its physical value. In \cite{Aoki:2006br} we had carried out zero
temperature simulations at four different points with nonphysical light quark
masses at each lattice spacing and made an extrapolation down to the physical
point. It is hard to estimate the systematic errors of such extrapolations.
Obviously such errors might also influence the determination of our LCP. In
this paper we will use only the LCP determined using extrapolations in
\cite{Aoki:2006br}. In order to check the size of the systematics of these
chiral extrapolations, we decided to carry out new simulations \emph{directly
at the physical point} for the same lattice spacings as in \cite{Aoki:2006br}.
As it will be shown our approach of \cite{Aoki:2006br} was very accurate.

\subsection{Action, algorithm}

The lattice action is the same as we used in \cite{Aoki:2006br}. On the
algorithmic side we have made couple of improvements. We use Omelyan
integration scheme \cite{Takaishi:2005tz} to integrate the evolution equations
of Rational Hybrid Monte Carlo (RHMC) (for details on the RHMC algorithm see
\cite{Clark:2006fx}). The smallest two poles of the rational approximation of
the light quark determinant are put to a larger integration timescale, than the
remaining ones. The solver residual is set to $\epsilon_{\rm ff}=10^{-5}$, when
calculating the fermion force in the RHMC, and $\epsilon_{\rm act}=10^{-8}$ in
the RHMC action. The code works mostly in float precision, while smaller than $10^{-6}$ precisions
are reached by using mixed precision inverters.  The updates of the links and
momenta are done in very large precision (80-bit or more), which results in an
exactly reversible algorithm. The reversibility is thus not effected by the
tolerance of the fermion force solver ($\epsilon_{\rm ff}$).

Our code is ported to two types of architectures: Intel PC equipped with
Graphical Processing Units (see \cite{Egri:2006zm}) and BlueGene/P. 

\subsection{Simulation points}
In Table \ref{ta:zero} we give the lattice spacings and the number of
trajectories for our zero temperature ensembles. These runs are done at the
physical values of the light and strange quark masses. We also show the quark masses of our old runs, 
which were used to carry out the chiral extrapolations to the physical point.
\TABULAR{|c|c|c|c|c|}
{
\hline
$\beta$ & $N_t\times N_s^3$ & \# traj & $m_{ud}/m_{ud}^{\rm LCP}$ & $m_{ud}/m_{ud}^{\rm LCP}$ in \cite{Aoki:2006br} \\
\hline
3.45 & $32\times 24^3$   &  1500 & 1 & 3, 5, 7, 9\\
3.55 & $32\times 24^3$   &  3000 & 1 & 3.5, 5, 7, 9\\
3.67 & $48\times 32^3$   &  1500 & 1 & 4, 6, 7.5, 9.5\\
3.75 & $48\times 40^3$   &  1500 & 1 & 4, 6, 8, 10\\
3.85 & $64\times 48^3$   &  1500 & 1 & --\\
\hline
}
{\label{ta:zero} Gauge coupling, lattice size, number of trajectories for our
zero temperature simulation points. The light and strange quark masses are set
to their physical values, ie. they are on the LCP as described in the text.
Next column shows, which light quark masses were used in \cite{Aoki:2006br} to
carry out the chiral extrapolations.}
The lattice volumes were chosen so that the continuum finite volume 
corrections
were below 0.5\% for the pion and kaon masses and decay constants
\cite{Colangelo:2005gd}. We measured gauge observables, chiral condensates and
susceptibilities after every, and hadron correlators after every tenth
trajectory. We performed correlated fits with the hadron propagators by using
the appropriate formulas for staggered mesons and baryons as described in
\cite{Ishizuka:1993mt}.  When extracting nucleon masses, we observed similar
ambiguities when using different quark sources as described in
\cite{Bernard:2001av}. We decided not to use them in the further analysis.

\subsection{Checking chiral extrapolations}
\label{ss:check}
First let us take a look at the pion and kaon masses (see Figure
\ref{fi:xtrap}). In \cite{Aoki:2006br} we used different fit formulas to
extrapolate to the physical point: for the kaon mass square the fit function was
linear in the quark mass, for the pion it was cubic. For the decay constants we
used a linear function plus a logarithmic $m_{ud}\log m_{ud}$ term with
unconstrained coefficients. Comparing the chiral extrapolations with results of
the direct simulations we find a remarkable agreement. For all four quantities
the difference is on the 1\% level for all lattice spacings.
\FIGURE[t]{
\includegraphics*[width=14cm,bb=18 434 591 718]{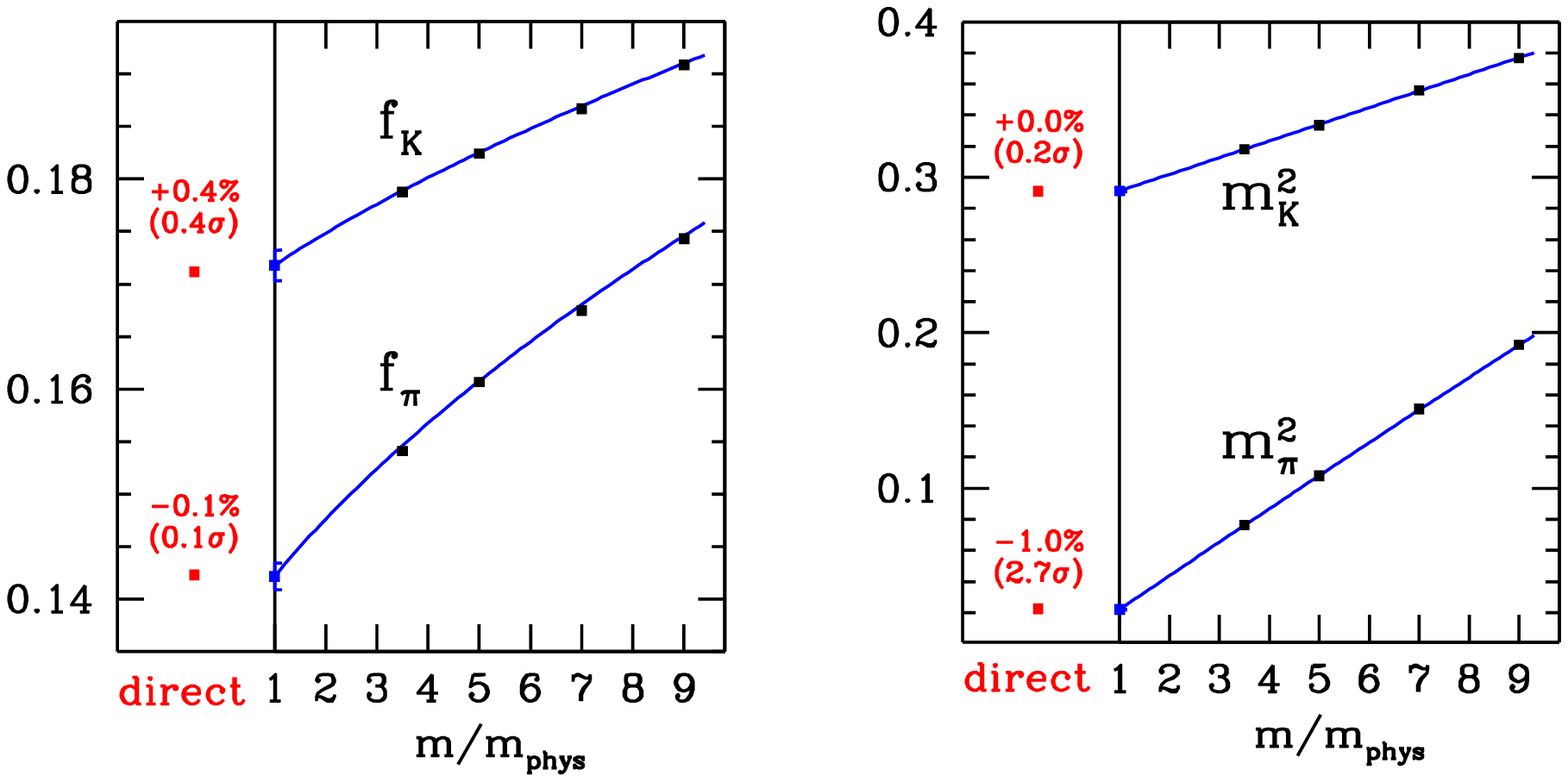}
\caption{\label{fi:xtrap} Chiral extrapolation vs. direct simulation of
the pseudoscalar decay constants and masses for $\beta=3.55$. 
Since this point has the highest statistics, any mismatch between the 
extrapolation and the direct result would be most pronounced here. We
do not observe such a mismatch.
Black points are data
from \cite{Aoki:2006br}, blue lines are our fit functions also from
\cite{Aoki:2006br}, which were used to extrapolate to the physical point, red
points are the results of the new simulations at the physical point. All values
are in lattice units.} }

We have also studied the effect of our extrapolations in case of the additive
renormalization constant of the chiral susceptibility. 
One expects that a slight change in the additive constant does not change
the position of a peak and, indeed the uncertainty of the 
extrapolation turned out to be
negligible on the location of the transition temperature (see the finite
temperature section).

\subsection{Setting the scale}
\label{ss:scale}
\TABULAR{|c|c|c|c|c|c|}
{
\hline
$\beta$ & 3.45 & 3.55 & 3.67 & 3.75 & 3.85 \\
\hline
$a(m_\pi)$[fm] & 0.2832(2) & 0.2193(1) & 0.1548(2) & 0.1267(2) & 0.1002(1) \\
$a(m_K)$[fm] & 0.2782(2) & 0.2153(1) & 0.1524(1) & 0.1246(1) & 0.0991(1) \\
$a(f_K)$[fm] & 0.286(2) & 0.217(1) & 0.153(1) & 0.123(1) & 0.097(1) \\
\hline
$a({\rm avg})$[fm] & 0.2824(6) & 0.2173(4) & 0.1535(3) & 0.1249(3) & 0.0989(2) \\
\hline
err[\%] & 1.5 & 0.9 & 0.9 & 1.4 & 1.5 \\
\hline
}
{\label{ta:afm} Lattice spacings obtained from different quantities (pion and
kaon masses and the kaon decay constant as well as the average of the three ).
Errors in parentheses are the quadratic sum of statistical and -- in case of $f_K$ --
experimental errors. The last row shows the maximum deviation from the average spacing, which we
consider as the systematic error of our scale setting.}

In \cite{Aoki:2006br} we have determined the Lines of Constant Physics and the
scale using three quantities: kaon and pion mass and kaon decay constant. There
we were using chiral extrapolations.  Now we can check directly at the physical
point, how consistent are the scales obtained from these three quantities (see
Table \ref{ta:afm}). We take $m_\pi=135$ MeV, $m_K=495$ MeV and $f_K=155.5$ MeV
for the physical values \cite{Amsler:2008zz}\footnote{In \cite{Aoki:2006br} we
used the 
Particle Data Group \cite{Eidelman:2004wy} value of $f_K=159.8$ MeV.
Note, however, that in the last 2.5 years the
Particle Data Group has reduced the central value of $f_K$ by about 3\%,
which \cite{Amsler:2008zz} reduces our $T_c$ values in physical units by 
the same amount.}. 
If the determination of
the LCP in \cite{Aoki:2006br} were completely correct, then the three
different quantities would give the same lattice spacing.
As it can be seen 
the deviation from the average of the three scales is always less
than 2\%. In \cite{Aoki:2006br} we have claimed a 2\% uncertainty in the scale
setting, so our current findings completely justify the previous results. 

We will use this average scale in our finite temperature analysis and consider
this 2\% as an uncertainty of the transition temperature arising from the zero
temperature simulations. 
\FIGURE[t]{
\includegraphics*[width=7cm,bb=18 434 290 695]{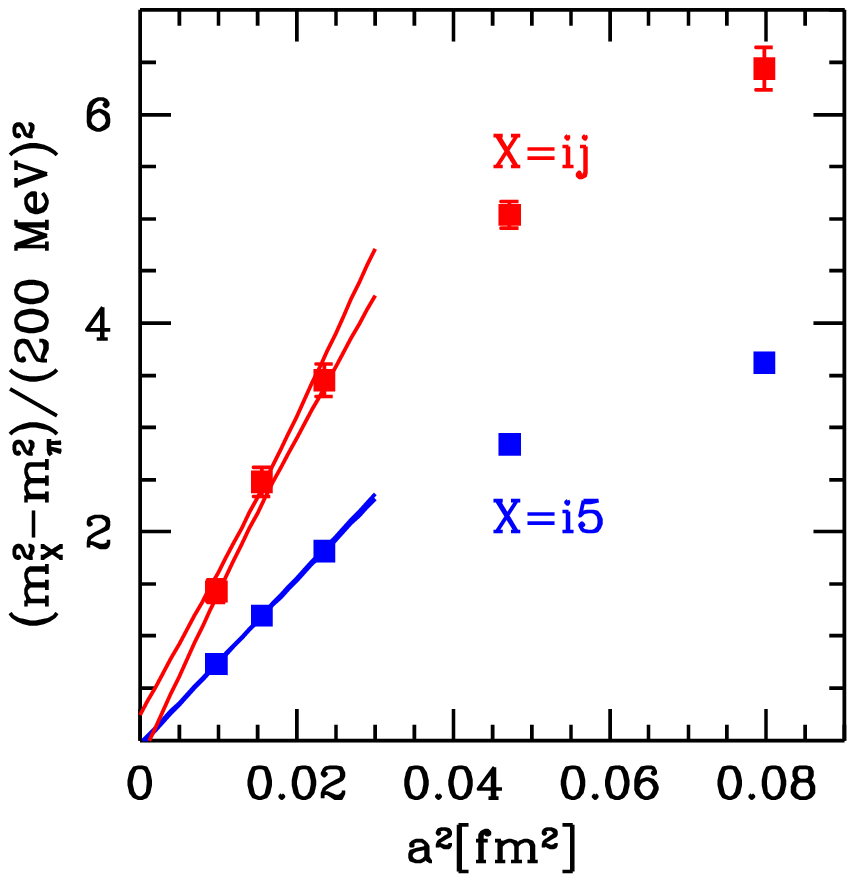}
\caption{\label{fi:split} 
Mass squared difference of the non-Goldstone pions (i5 and ij) and the Goldstone pion
as a function of the lattice spacing squared.
}
}

In the following subsections we will present some results for zero temperature
observables: hadron and quark masses and decay constants. In these cases we
attempt to eliminate even this small 2\% systematic error. 
On the ensembles of Table \ref{ta:zero} in addition to our
measurements we measure propagators, where the quark masses
are set to $\pm20$\% of the physical strange quark mass and $\pm10$\% of the
physical light quark mass. By interpolating between these quark mass values we
look for those strange and light quark mass parameters, where $m_\pi/f_K$ and
$m_K/f_K$ take their experimental values exactly. The so obtained correction to
the quark masses has turned out to be always less than 7\%. At this corrected
point we measure the ratios of various observables. This procedure takes
into account only the change in the operator due to the variation in the quark
mass, the slight change in the background gauge field is neglected. However,
as we checked it for a few points,  in ratios of observables this effect 
largely cancels and the uncertainty related to this procedure remains far
below our statistical accuracy.

\subsection{Taste violation} 

The taste symmetry breaking of the staggered fermion discretization splits up
the originally degenerate masses of the pion multiplet, leaving only one pion
massless in the chiral limit. Taste symmetry violation has to vanish in the
continuum limit, otherwise the staggered discretization would fail to be a
proper fermion discretization.  Therefore, it is important to check whether the
pion splitting vanishes when carrying out a continuum extrapolation using the
available lattice spacings.  This extrapolation provides a useful
hint where the scaling regime is expected to start. We take two
representatives of the non-Goldstone pions: i5/MVII and ij/MVIII (the notations
are that of MILC and \cite{Ishizuka:1993mt}). Let us take a look at the quadratic
mass difference of the non-Goldstone pions and the Goldstone pion as a
function of the lattice spacing squared (see Figure \ref{fi:split}). One can
clearly see that the taste violation decreases with decreasing lattice spacing.
Moreover we can also observe that lattice spacings which are larger than $a\sim
0.15 {\rm fm}$ (the corresponding critical temperature in lattice units is $1/N_t\sim
1/8$) are not in the $a^2$-scaling regime in the case of these
quantities. The taste violation for the three finest lattice spacings can be
extrapolated to zero lattice spacing: for both type of non-Goldstone pions the
splitting is consistent with zero in the continuum limit.  

\subsection{Hadron masses, $m_s/m_{ud}$ and $f_K/f_\pi$}
A necessary condition for the correctness of the finite temperature results is
that zero temperature observables in the continuum limit are consistent with
experiments. Moreover, the lattice spacing dependence of the zero temperature
observables can give a hint on the lattice spacing range, where lattice
artefacts are expected to scale as $a^2$. 

Let us first take a look at various hadron masses (see left panel of Figure
\ref{fi:spect}). At the top of the figure the mass of the $\Omega$ baryon is
plotted as a function of the lattice spacing squared. The red band is the
experimental value of the $\Omega$ mass together with its uncertainty (to which
the experimental uncertainty of our scale fixing quantity $f_K$ also
contributes). Our four finest lattice spacings are nicely consistent with the
experiments. This fact confirms the correctness of the $f_K$-based scale setting
procedure. 
In other words, we have shown that performing the scale setting with the
$\Omega$ mass would give the same continuum values for $T_c$ in physical
units.

The $\phi(1020)$ meson mass is plotted in the middle. The open and solid
symbols correspond to two different vector meson operators (MIII and MIV using
the notations of \cite{Ishizuka:1993mt}), they are supposed to give the same mass
in the continuum limit. We use only the connected part of the operators when
evaluating the propagators 
(the disconnected part is very expensive to calculate; however, as large 
scale T=0 simulations show \cite{Aoki:2008sm}, omitting the disconnected 
part for $\phi(1020)$ could provide the proper scale, the uncertainty related to this 
choice is subdominant). 
The plot shows also an agreement with the experiment (red band).
\FIGURE[t]{
\includegraphics*[width=14cm,bb=32 405 580 702]{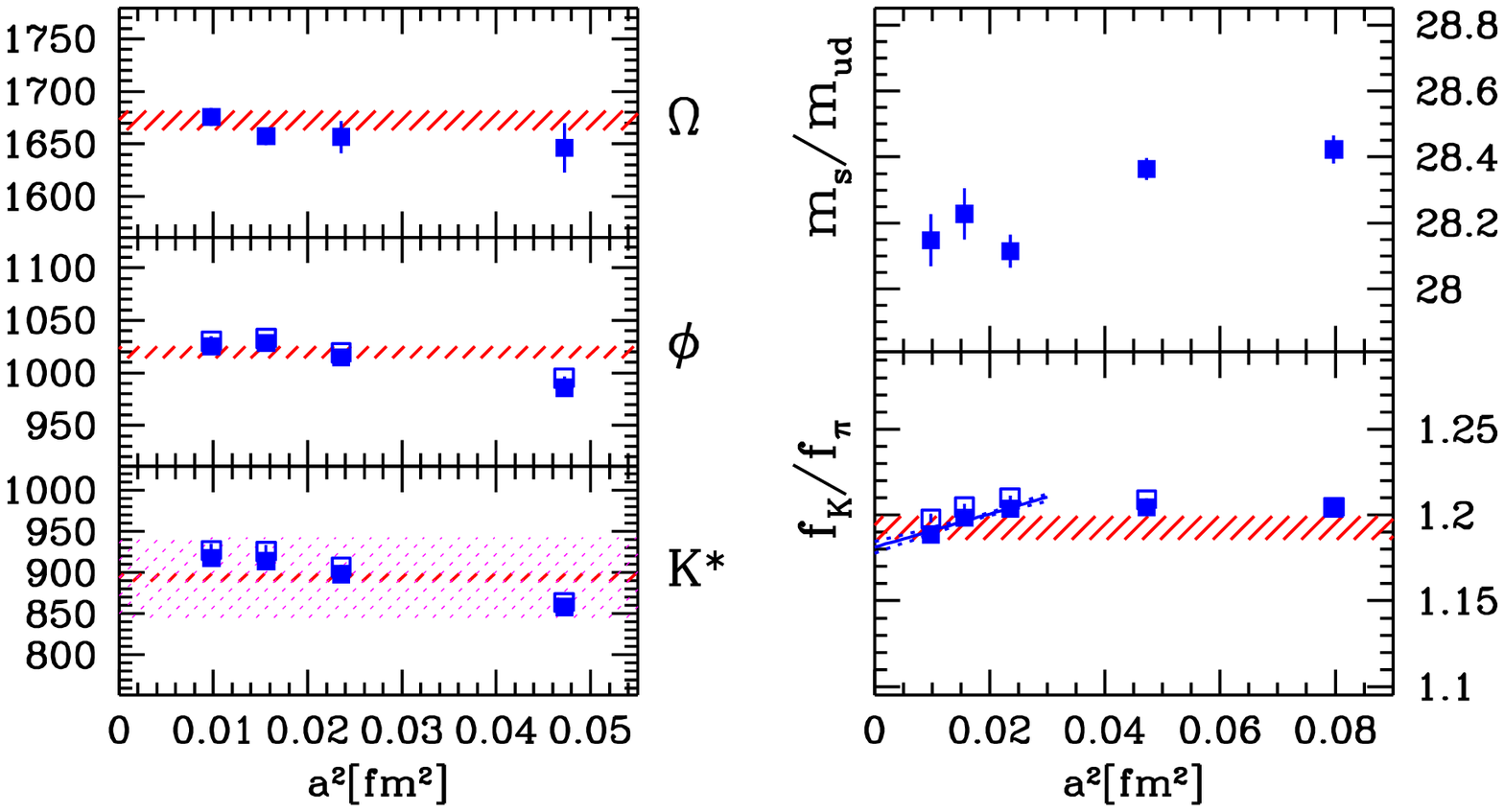}
\caption{\label{fi:spect}
Left panel: masses of $\Omega$ baryon, $\phi(1020)$ meson and $K^*(892)$ meson in MeV 
on our four finest lattices as a
function of the lattice spacing squared. Right panel: quark mass ratio and $f_K/f_\pi$ for
all five ensembles. See text for a detailed explanation.
}
}

The lower plot shows the $K^*(892)$ vector meson mass. Open and solid symbols
are the two vector meson operators, as in the case of $\phi(1020)$. The
agreement is somewhat worse than for the other two masses. However one has to
keep in mind that at the physical point in our boxes the strong decay of
$K^*(892)$ is kinematically allowed. Our operators are supposed to have negligibly
small coupling to scattering states and couple mostly to the resonance. The
resonance energy level at a given volume is not necessarily the central value
of the resonance ($m_{K*}$), but it might be some other value within the
resonance distribution (which has $\Gamma_{K*}$ width). Therefore, beside the
red band, which is the experimental value of the $K^*(892)$ mass, we also draw
a $2\Gamma_{K*}$ wide magenta band inside which the resonance levels are expected to
appear.

The right panel of Figure \ref{fi:spect} shows the ratio of the 
strange and light
quark masses. Note, that this is not the ratio along the LCP (which was fixed
to $m_s^{\rm LCP}/m_{ud}^{\rm LCP}=27.3$), but the ratio of the quark masses
after carrying out the correction to the LCP as described in Subsection
\ref{ss:scale}. As one can clearly see there is no observable lattice spacing
dependence for our three smallest lattice spacings. Therefore it is completely
justified to take the result on the finest lattice spacing as the continuum
estimate for the quark mass ratio:
$m_s/m_{ud}=28.15.$
The statistical error is on the 0.4\% level, the systematic uncertainties
are somewhat larger. 

On the lower part of the right panel we plot the ratio of kaon and pion decay
constants against the lattice spacing squared for all five ensembles. The red band
is the current best estimate for $f_K/f_\pi$ including the uncertainty. Opened
symbols are the original lattice data, whereas the solid ones contain the
continuum limit finite volume corrections \cite{Colangelo:2005gd}. For the
three finest lattice spacings we can observe a clear decreasing tendency. An
extrapolation with an $a^2$ scaling function yields
$f_K/f_\pi=1.181$
in the continuum limit. 
The statistical error of $f_K/f_\pi$ is on the 0.3\% level. The systematic 
uncertainties are of the same order of magnitude. 

A detailed analysis of the systematic uncertainties of $m_s/m_{ud}$ as well as
$f_K/f_\pi$ is quite interesting from the T=0 physics point of view and will be
published elsewhere~\cite{Aoki:2099}. In this forthcoming publication we
discuss the masses of the $\Omega$ baryon, the $K^*(892)$ meson and the 
$\phi(1020)$ meson in detail, too.

The basic message of this subsection can be summarized as follows. Using an
$f_K$ based scale setting procedure (see Subsection \ref{ss:scale}), the masses
of $\Omega$, $K^*(892)$, $\phi(1020)$ and the pion decay constant are
consistent with their experimental values on our finest lattices. This implies
that independently of which of these quantities is used for scale setting, we
would obtain the same results in the continuum limit.

\subsection{Static quark potential} 
\label{ss:r0}
\FIGURE[t]{
\includegraphics*[width=14cm,bb=18 434 591 718]{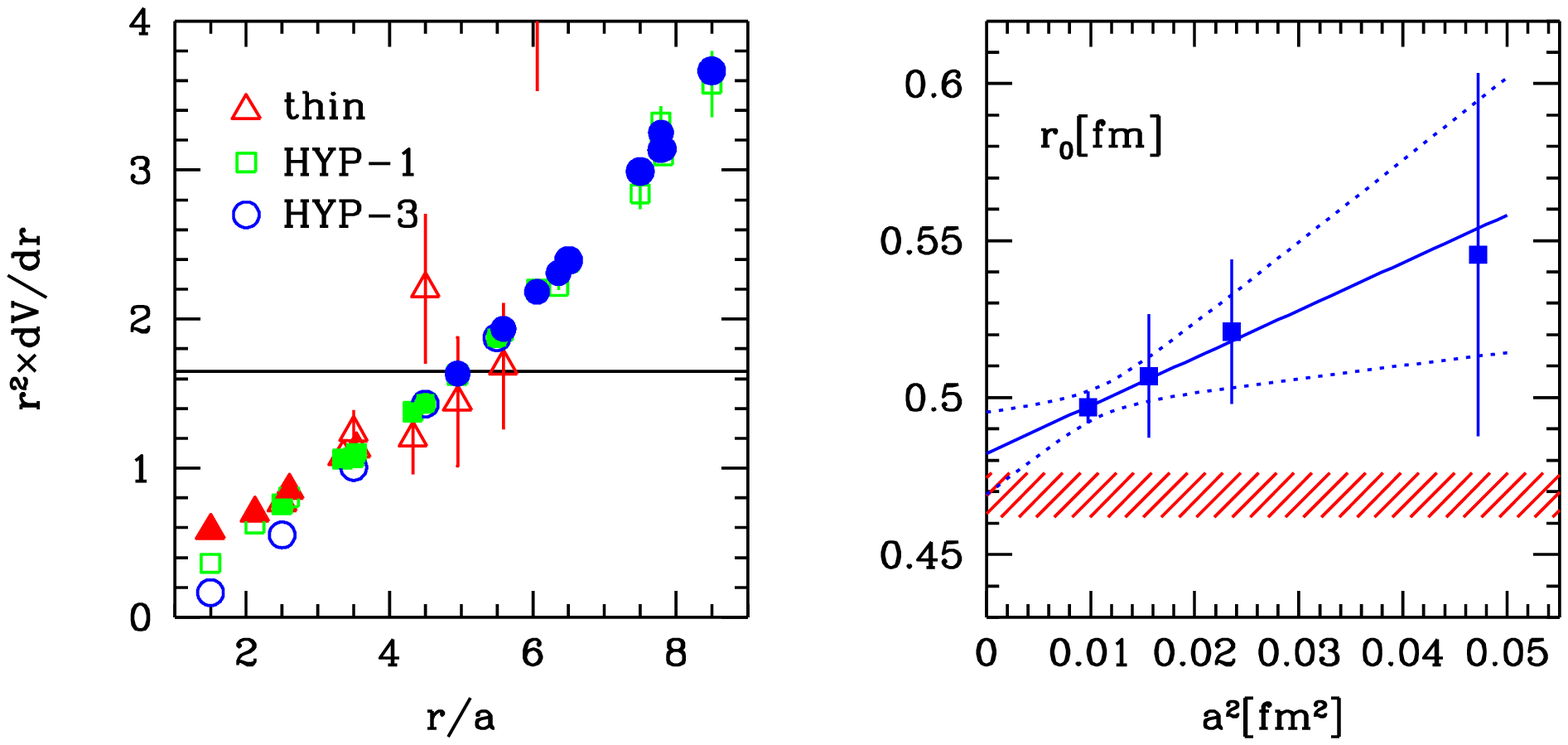}
\caption{\label{fi:r0}Left panel:
the static quark force multiplied by the distance squared for three 
different smearing levels. The horizontal line corresponds to $1.65$, which
value defines the Sommer-scale.
Right panel: Sommer-scale in physical units as a function of the
lattice spacing squared. The red band is the $r_0$ determination from \cite{Gray:2005ur}.
}
}

A popular way to fix the scale in lattice QCD is to use quantities related to
the static quark potential $V(r)$, like the string tension or Sommer scale
\cite{Sommer:1993ce}.  The major advantage compared to other methods
is that there are no ambiguities in the construction of operators due to
staggered taste violation, since the Wilson-loops are built up only
from the 
gauge fields. A disadvantage is that on coarse lattices (which are usual
in thermodynamical calculations) the static quark potential determination is
burdened by sizeable systematics. It is hard to extract ground state energy
levels of the static quark-antiquark pair (compared to mass extraction in
hadron spectroscopy), since the signal disappears quickly in the noise.

We use the following gauge link smearing recipe to increase our signal/noise
ratio. The spatial links are smeared by 30 steps of APE smearing
\cite{Albanese:1987ds}, this reduces the excited state contamination while
keeping the ground state energy intact for all distances. We also smear the
timelike links by 3 steps of HYP smearing \cite{Hasenfratz:2001hp}, keeping all
the intermediate steps, too. This decreases the noise substantially, however
distorts the potential for small distances. By comparing the results of zero,
one, two and three steps of HYP smearing we can determine a minimal distance
for each level of HYP smearing steps, above which that smearing level can be
safely used, ie. there is no significant distortion in the potential.  Let us
illustrate this on the left panel of Figure \ref{fi:r0}, where the quantity $r^2dV/dr$
is plotted as a function of the distance for our finest lattice
spacing ($\beta=3.85$). Different symbols are used for the different
HYP-smearing levels. The filled symbols indicate which smearing level was used
at a given distance. For small distances the smearing distorts the potential,
there we use no smearing at all.  As the distance increases, the distortion
effect becomes gradually smaller, which makes it possible to use higher smearing
levels. 

The Sommer scale ($r_0$) is defined as the distance where $r^2dV/dr=1.65$.  We 
estimate the systematic errors as follows: beside the potential we make fits to the
force itself, we consider different interpolating functions and different types
of Wilson-loops. For our two coarsest lattice spacings these systematics turned out to
be large.  We measure therefore the $r_2$ scale, which is defined as the point where
$r^2dV/dr=2$. On coarse lattices it has considerably smaller systematic
errors than what $r_0$ has. On the right panel of Figure \ref{fi:r0} we show the
lattice spacing dependence of $r_0$, on the coarsest lattices its value was
derived from that of $r_2$. A clear downward trend can be observed as the lattice spacing is decreased,
in the continuum limit we get
\begin{eqnarray}
r_0=0.48(1)(1) {\rm ~fm.}
\end{eqnarray}

The first error comes from the statistical and systematic error of the $r_0$
determination, whereas the second is from the uncertainty of the scale
determination.  This is consistent with an other staggered $r_0$ determination
\cite{Gray:2005ur}: $r_0=0.469(7)$ fm, which is the value used by the 'hotQCD'
collaboration in their thermodynamical studies. Let us mention here that there
are other $r_0$ determinations in the literature: $0.467(33)$~fm from the QCDSF
collaboration~\cite{Gockeler:2005rv} and $0.492(6)(7)$ fm from 
PACS-CS~\cite{Aoki:2008sm}. The differences between the results suggest
the possibility that the systematic errors are underestimated in the $r_0$
determination.

\section{Finite temperature simulations}
\label{se:fint}

In \cite{Aoki:2006br} we used four lattice spacings, $N_t=4, 6, 8$ and $10$ to
study the lattice spacing dependence of thermodynamical observables.  The
quark masses were set to their physical values, i.e. to $m_{ud}^{\rm LCP}$ and
$m_{s}^{\rm LCP}$. In case of the transition temperatures we carried out a
continuum extrapolation based on the finest three lattices ($N_t=6,8$ and
$10$). 

In this work we extend our finite temperature data set by simulations on
$N_t=12$ and $16$ lattices with physical quark masses. 
As we have shown with our finite volume analysis \cite{Aoki:2006we}
the temperature dependence changes only very little in the $N_s/N_t$=3--5
range. Therefore,
we generated between
1500 and 3500 trajectories on $12\cdot36^3$ lattices at 18 different
temperature values and on a $16\cdot48^3$ lattice at one temperature.  The
lattice scale range which we examined in Section \ref{se:zerot}, covers nicely
the transition regime of the $N_t\le 12$ lattices. In case of the strange quark
number susceptibility we will show results for somewhat higher temperatures
($>210$ MeV on $N_t=12$ and $260$ MeV on $N_t=16$). In this case the scale was
determined by a method which will be published elsewhere \cite{Borsanyi:2099}.

In the following we present the results and compare them with those of the
'hotQCD' collaboration.

\subsection{Renormalized chiral susceptibility}
\FIGURE[t]{
\includegraphics*[width=14cm,bb=18 434 591 718]{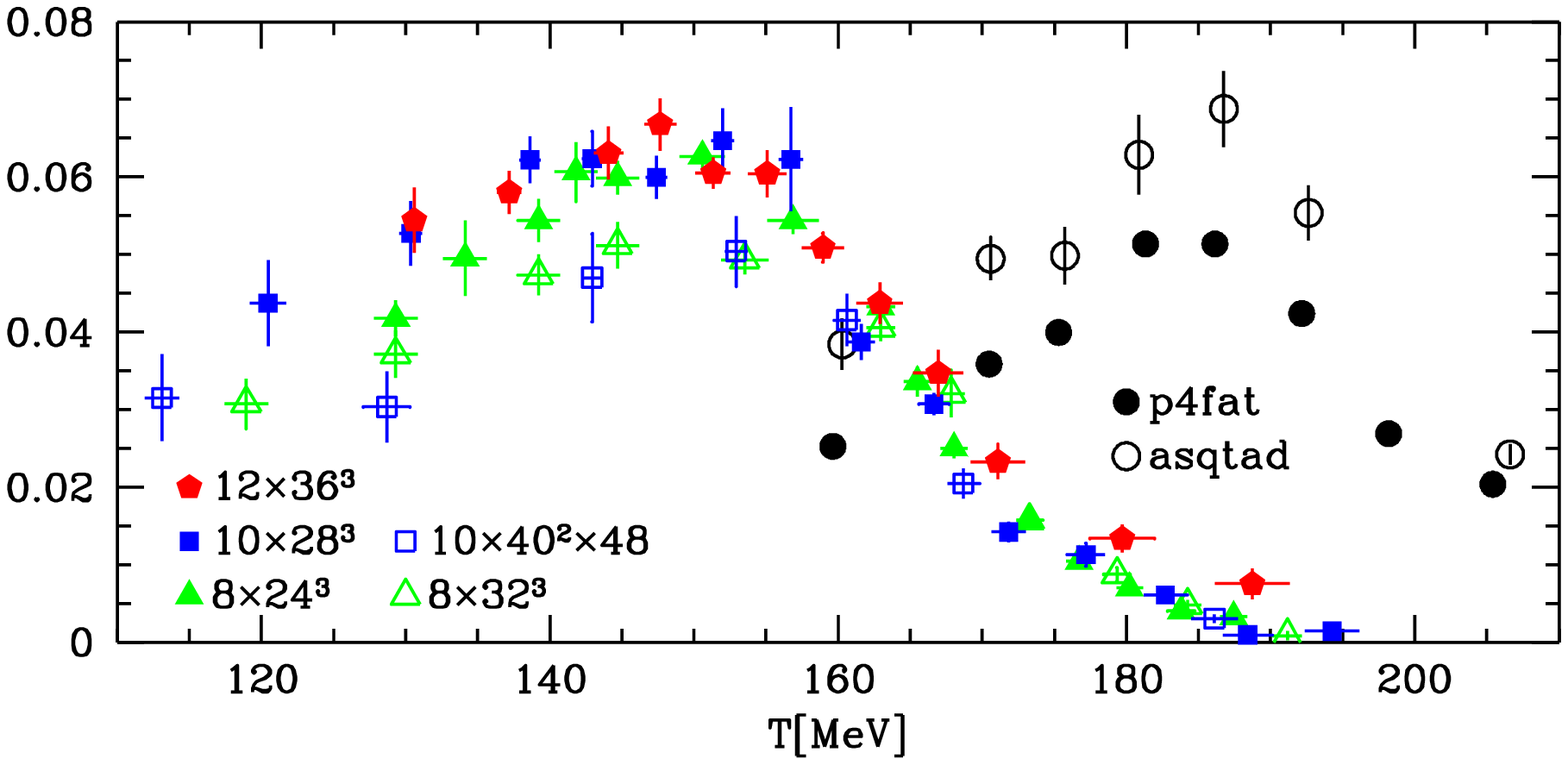}
\caption{\label{fi:susc}
Renormalized chiral susceptibility normalized by $T^4$. Open colored symbols
are results on smaller volumes (with aspect ratio $N_s/N_t$ around 3), whereas
filled colored symbols are results on larger volumes (with aspect ratio four).
For comparison results of the 'hotQCD' collaboration with two different fermion
actions on $N_t=8$ 
are also shown, they have been rescaled by an appropriate factor (see
text).
}
}

The light quark chiral susceptibility ($\chi_{\bar{\psi}\psi}$) is minus 
one times the second
derivative of the free energy density with respect to the light quark
mass. It is ultraviolet divergent. In \cite{Aoki:2006we} we proposed the
following renormalization recipe. Since the ultraviolet divergences are
independent of the temperature, subtracting the susceptibility at zero
temperature from the susceptibility at finite temperature removes the additive
divergences:
\begin{eqnarray}
\Delta \chi_{\bar{\psi}\psi}= \chi_{\bar{\psi}\psi}(T)-\chi_{\bar{\psi}\psi}(T=0).
\end{eqnarray}
The multiplicative renormalization can be done by multiplying 
by the square of the bare quark mass:
\begin{eqnarray}
\Delta \chi_{\bar{\psi}\psi} \to m_{ud}^2\cdot\Delta \chi_{\bar{\psi}\psi}.
\end{eqnarray}

On Figure \ref{fi:susc} we plot this renormalized chiral susceptibility
normalized by $T^4$ as a function of the temperature.  We show results for
three different lattice spacings ($N_t=8,10$ and $12$). In case of $N_t=8$ and
$10$ we have the results on two different volumes as well,  the larger volumes
are plotted with filled symbols. The finite temperature data on $N_t=8$ and
$10$ was taken from our old paper. The renormalization was carried out with the
new zero temperature results (see Subsection \ref{ss:check}). The scale has
also slightly changed due to the change in the experimental value of the $f_K$
in the Particle Data Group (see Subsection \ref{ss:scale}). This results in an
overall $\sim 5$ MeV downward shift in the temperature compared to what we
reported in \cite{Aoki:2006br}. 

We see no considerable lattice artefacts, in particular the new $N_t=12$ results are 
consistent with the $N_t=10$ ones from our old data set.
A small volume dependence can be seen in the height of the susceptibility peak,
but the volume dependences of the width and the position are not significant within
the present statistics. 

In order to help comparisons with other approaches we also provide the
temperature dependence for the renormalized chiral susceptibility normalized by
$T^2$ or not normalized by any power of $T$, at all (see Figure
\ref{chi_norm}).  As it can be seen the curves are gradually shifted to the
right, resulting in increasing transition temperatures defined from the peak positions (see
Table \ref{ta:tc_results}). This is a feature of the crossover type transition,
different definitions generally result in different temperature values.
\FIGURE[t]{
\includegraphics*[width=14cm,bb=18 434 591 718]{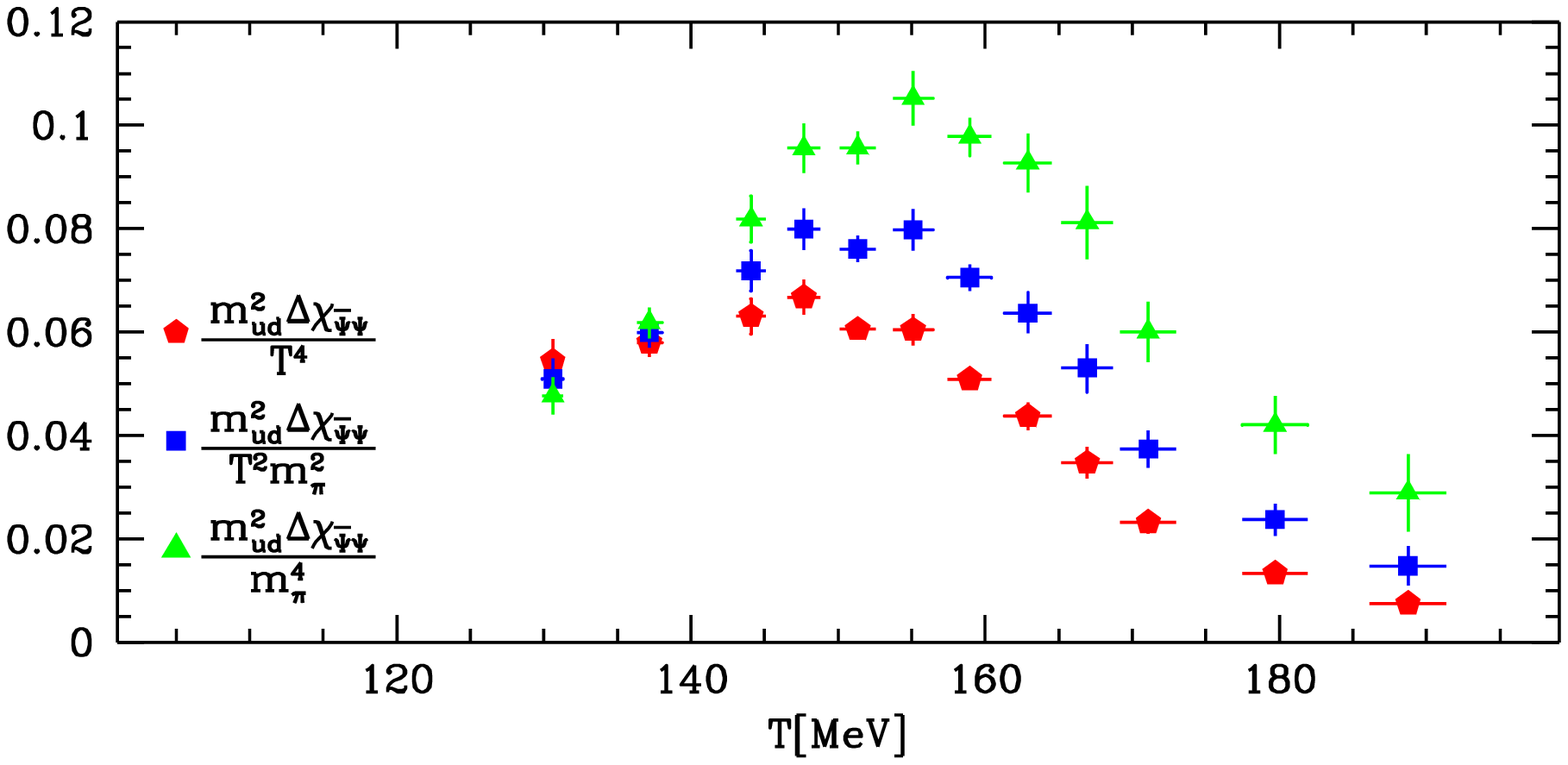}
\caption{\label{chi_norm}
Comparison of the temperature dependence of the renormalized
chiral susceptibility normalized by various powers of $T$. Only
our $N_t=12$ data are shown. Different symbols correspond to different 
normalizations.}
}

Now let us make the comparison with the results of the 'hotQCD' collaboration.
First let us consider the data of \cite{Detar:2007as}, which uses 'asqtad'
fermion discretization. The light quark masses in our simulations and in the
simulations of~\cite{Detar:2007as} are quite different. 
The latter uses three times
larger light quark masses than the physical, which is used in our work. Since
the renormalized chiral susceptibility depends strongly on the quark mass,
there is no problem with the fact that the height of the susceptibility is
considerably larger in the simulations of \cite{Detar:2007as} than what we
obtain. For convenience we multiply the results of \cite{Detar:2007as} by a
factor of 0.4, these points are the black filled circles on Figure \ref{fi:susc}.
We also plot the data obtained using the 'p4fat' action \cite{Soltz:2008} (black
opened circles). These results were multiplied with a factor of 0.15 for similar
reasons as for the 'asqtad' action. Both 'hotQCD' results were simulated on
$N_t=8$ lattices. Whereas the results of the 'hotQCD' group agree on the
position of the susceptibility peak, we observe a huge disagreement with our
data, which is in the order of $35$ MeV.  It is unclear whether
an effect of this
size can be explained only by the difference in the quark masses. Most probably
the origin is somewhere else: as we will see soon, much less quark mass
dependent quantities also show similar discrepancies.

\subsection{Renormalized chiral condensate}
\FIGURE[t]{
\includegraphics*[width=14cm,bb=18 434 591 718]{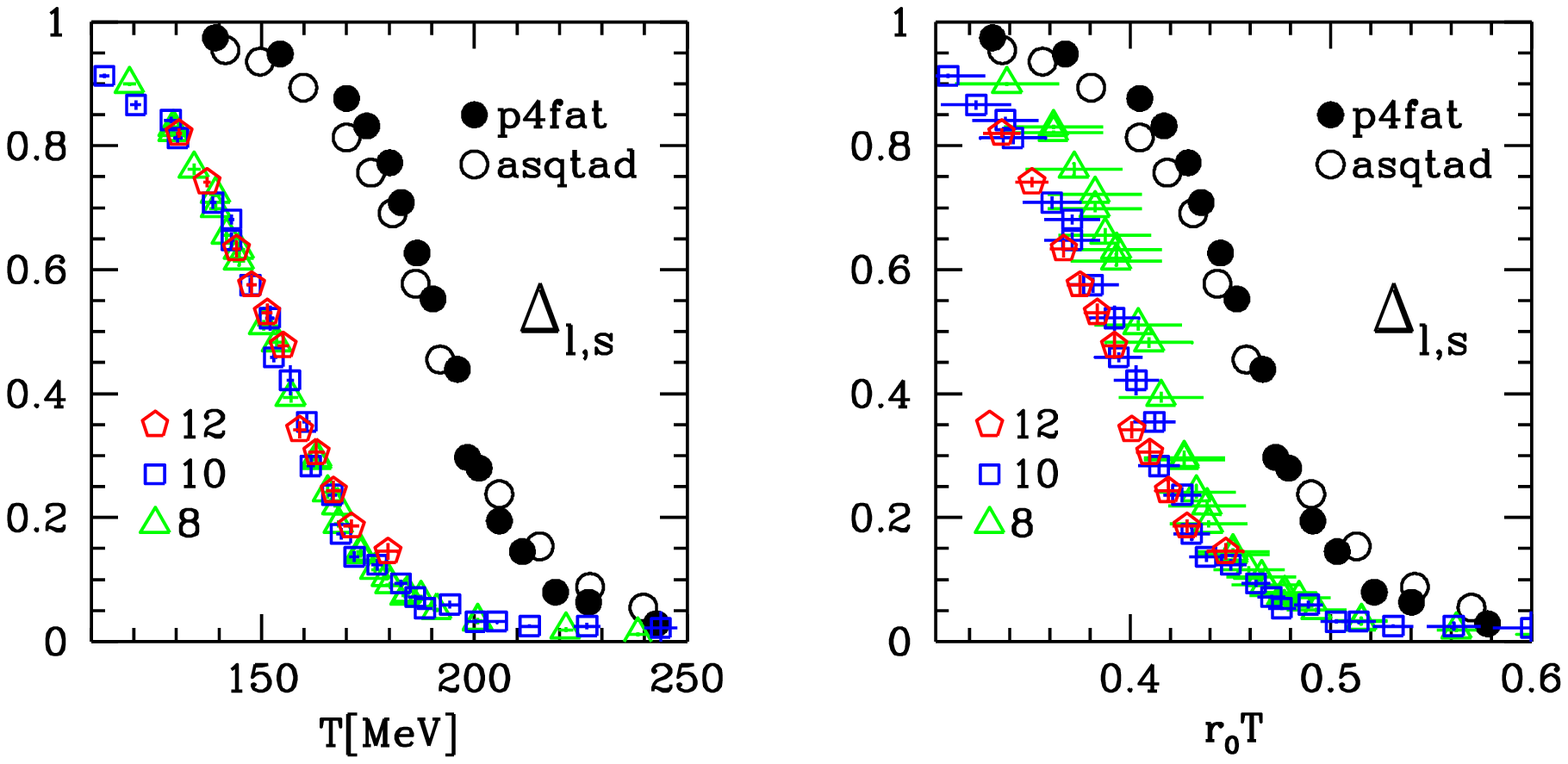}
\caption{\label{fi:pbp}
Renormalized chiral condensate as a function of the temperature.  On the left
panel the temperature is given in physical units, whereas on the right in the
units of the Sommer scale ($r_0$).  Colored opened symbols are the 
results on
$N_t=8,10$ and $12$ lattices. For comparison results of the 'hotQCD'
collaboration with two different fermion actions on $N_t=8$ are also shown.
}
}
The light quark chiral condensate ($\langle \bar{\psi}\psi \rangle$) is minus 
one times the
first derivative of the free energy density with respect to the light
quark mass. It is ultraviolet divergent, a possible way of removing divergences
was proposed in \cite{Cheng:2007jq}.  \emph{If one assumes} that the additive divergences of
the free energy density depend on the quark masses only through the
combination $m_{ud}^2+m_s^2$, then one can get rid of the additive divergences in $\langle \bar{\psi}\psi \rangle$
by using the strange quark condensate ($\langle \bar{s}s \rangle$):
\begin{eqnarray}
\Delta_{l,s}= \langle \bar{\psi}\psi \rangle-\frac{2m_{ud}}{m_s}\langle \bar{s}s \rangle.
\end{eqnarray}
The remaining multiplicative divergences can be removed by dividing with the same quantity at 
zero temperature:
\begin{eqnarray}
\Delta_{l,s}\to \frac{\Delta_{l,s}(T)}{\Delta_{l,s}(T=0)}.
\end{eqnarray}

On Figure \ref{fi:pbp} we plot this quantity as a function of the
temperature. There is no significant lattice spacing or volume dependence for
lattices of $N_t=8,10$ and $12$ and for aspect ratios 3-4. For comparison we take
the $N_t=8$ data of the 'hotQCD' collaboration from \cite{Karsch:2008fe}.
Similar to the case of the chiral susceptibility we find a huge disagreement
between the curves in the transition regime. Again the shift between the curves
of the different groups is in the order of $35$ MeV. 
\FIGURE[t]{
\includegraphics*[width=7cm,bb=18 434 290 695]{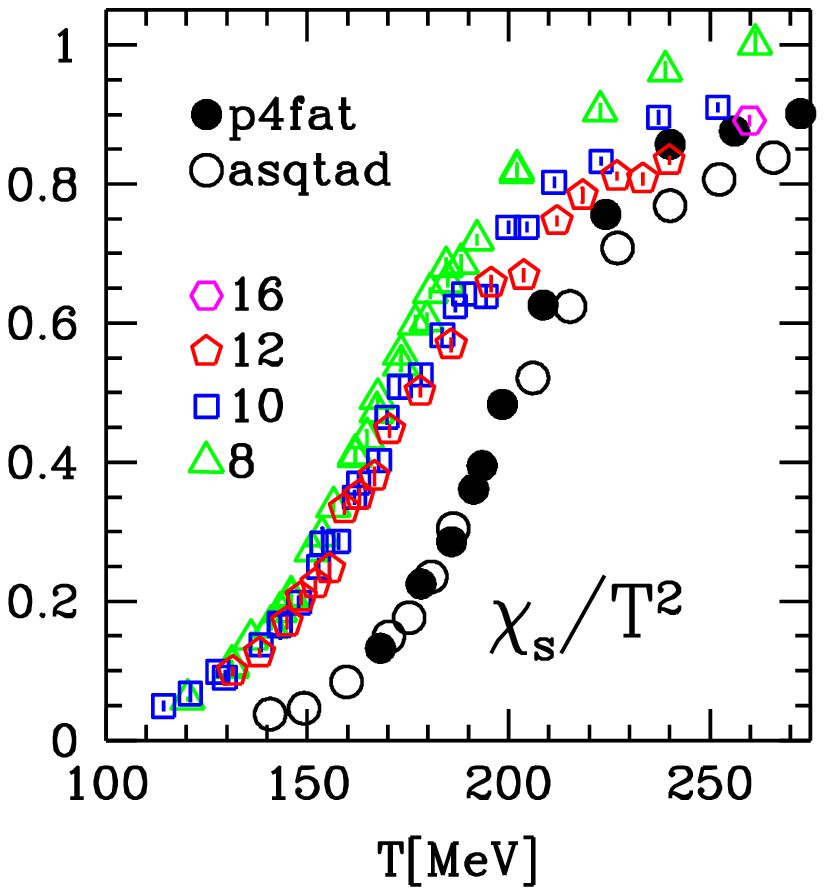}
\caption{\label{fi:qns}
Strange quark number susceptibility normalized by $T^2$. Colored opened symbols
are results on $N_t=8,10$ and $12$ lattices. We have an additional point on an
$N_t=16$ lattice at our highest temperature.  For comparison results of the
'hotQCD' collaboration with two different fermion actions on $N_t=8$ are also shown.
}
}

One might think that the different scale fixing methods used by the different
collaborations are responsible for this $35$ MeV discrepancy. The 'hotQCD'
collaboration uses the Sommer scale in their scale fixing procedure, so it can be
enlightening to look at our results, if the temperature is given in units of
the Sommer scale (right panel of Figure \ref{fi:pbp}). The scaling is somewhat
worse in terms of this quantity, however for the finest lattices the
discrepancy is still present. This does not come as a surprise, since the $r_0$
in physical units obtained in Subsection \ref{ss:r0} is perfectly consistent
with the one used by the 'hotQCD' group.

\subsection{Strange quark number susceptibility}
The strange quark number susceptibility ($\chi_s$) is defined as minus one times the derivative
of the free energy density with respect to the square of the strange
quark chemical potential. It is conveniently normalized by $T^2$, by which it
will asymptotically reach one as the temperature is increased to infinity (Stefan-Boltzmann limit).

Our results on $N_t=8,10$ and $12$ are shown in Figure \ref{fi:qns}. We
observed no volume dependence, therefore we use the same symbols for the two
different aspect ratios. There is no significant lattice spacing dependence for
temperatures smaller than $\sim 170$ MeV, whereas for higher temperatures the
lattice artefacts are somewhat larger. This is expected, since in the
Stefan-Boltzmann limit the lattice artefacts are known to be large for our
action. We also have an additional point on a very fine lattice ($N_t=16$) at a
high temperature.

The comparison with the results of the 'hotQCD' collaboration (see Reference
\cite{Karsch:2007dp}) brings us to a similar conclusion as for the other two
quantities that we have considered before. Around the transition point there
is an approximately $20$ MeV shift between the results of the two groups.
For larger than $\sim230$ MeV temperatures our finer lattices are in good
agreement with the 'hotQCD' results.

\subsection{Transition temperatures}
\TABULAR{|c|c|c|c|c|c|c|}
{
\hline
 & $\chi_{\bar{\psi}\psi}/T^4$ & $\chi_{\bar{\psi}\psi}/T^2$ & 
$\chi_{\bar{\psi}\psi}$ & $\Delta_{l,s}$ & L &  $\chi_s$\\
\hline
this work                          & 146(2)(3) & 152(3)(3) & 157(3)(3) & 155(2)(3) & 170(4)(3) & 169(3)(3) \\
our work '06 & 151(3)(3)  & - & - & - & 176(3)(4) & 175(2)(4)\\
RBCBC          & - & 192(4)(7) & - & - & 192(4)(7) & -\\
\hline
}
{
\label{ta:tc_results} 
Transition temperatures for different observables and in different works. See the text for explanation.
}

In this subsection we present our results for the transition temperatures
obtained from different quantities (see Table \ref{ta:tc_results}). The first
three columns contain the transition temperatures of the renormalized
chiral susceptibility, each of them normalized differently: with $T^4$, $T^2$
and without any power of $T$. The peak position was determined by fitting  a
quadratic curve to the points around the peak.  The first error comes from the
statistical errors and from the variation of the fit range, whereas the second
error arises from the accuracy of our scale determination.  As it can be
clearly seen and as it has been already shown before, different normalizations
yield significantly different peak positions. 

In the next three columns the transition temperatures from the inflection point
of the renormalized chiral condensate, renormalized Polyakov loop and the
strange quark number susceptibility are given. These inflection points were
obtained by fitting cubic polynomials to the data. Systematic errors were
estimated by the variation of the fit ranges. 

We have also measured the width of the transition for all these observables
(the definition can be found in \cite{Aoki:2006br}). It is found to be in the
25-30 MeV range in all cases. 

In the second line we provide our previously published results from 2006
\cite{Aoki:2006br}. Our lattice results are in complete agreement with our
earlier findings, the reason for the approximately 5~MeV shift to lower $T_c$
values is almost completely due to the change of the experimental value of
$f_K$ provided by the Particle Data Group ($155.5$ MeV \cite{Amsler:2008zz}
instead of $159.8$ MeV \cite{Eidelman:2004wy}). Without this change in the
input parameter the change of the $T_c$ values would be about or less than 1
MeV.

We also include into the table the combined physical quark mass and continuum
extrapolated estimates of the RBC-Bielefeld collaboration (RBCBC)
\cite{Cheng:2006qk}. The RBCBC did not use renormalized quantities, furthermore
the transition temperature related to the Polyakov loop is determined from the
peak position of the Polyakov-loop susceptibility, which is different from our
definition. These differences are expected to be small compared to the
statistical and systematic uncertainties. The discrepancy between the
temperature values of the two collaborations is worryingly large, as it was
already emphasized in the Introduction.

\section{Conclusions, outlook}

We have improved our previous calculations on the transition temperature
\cite{Aoki:2006br} by
three means. First of all, the simulations for our zero temperature analysis
have been done with the physical values of the quark masses. Secondly, we
extended our hadron spectrum, decay constants, quark mass and static quark
potential measurements.  As a third improvement we have decreased the lattice
spacing at finite temperature by simulating $N_t=12$ lattices (and $N_t$=16 at
one point). 

For the first time in the literature we performed both the $T=0$ and $T>0$
analyses by 
simulating directly with
physical quark masses. This procedure eliminates all
uncertainties related to the extrapolation to the physical masses.
The analysis confirms that the uncertainty of our scale
determination is less than about 2\%. Moreover, all spectral quantities are
consistent with experiments and/or previous lattice calculations. 
This indicates that the finite temperature results are independent of which
quantity ($\Omega$, $K^*$ or $\Phi$ mass, or the pion decay constant) 
we chose for scale setting.

At finite temperature we determined the temperature dependence of several renormalized 
quantities. As a generic feature of any crossover, the transition temperatures 
obtained from different quantities are different, they range from 146 to 170~MeV. 
We have to emphasize again that these numbers correspond to an infinite volume
system. As an exploratory study in quenched QCD shows~\cite{Bazavov:2007zz}, 
for the typical volumes and boundary conditions 
realized at heavy ion collisions, the transition temperatures can be 
up to 30~MeV higher than the infinite volume values presented here and usually 
in the literature.

The new results at finite temperature are in good agreement with our previous 
findings. Note, however, that in the last 2.5 years
Particle Data Group has reduced the central value of $f_K$ by about 3\%, 
which reduces our $T_c$ values in physical units by the same amount. 
The lattice spacings used in this work
are smaller than in any previous lattice study. As
a consequence, the lattice artefacts seem to be small, 
there are even quantities, where the artefacts are not significant at all. 

We have taken a closer look at the disagreement between the results of current
thermodynamical calculations. We see approximately $20-35$~MeV 
difference in the transition regime between our results and those of the 'hotQCD'
collaboration. This difference can be observed between the temperature 
dependence 
of the curves for all the quantities that we have compared: the light quark
chiral susceptibility, renormalized chiral condensate and the strange quark number
susceptibility. Finding the reason for this disagreement seems to be a task for
the future.

As a final remark we have to mention that the staggered formalism used 
in this work and all other large scale thermodynamics studies 
may suffer from theoretical
problems. To date it is not proven that the 
staggered formalism with 2+1 flavors
really describes QCD in the continuum limit.
Therefore it is desirable to also study QCD thermodynamics
with a theoretically firmly established (e.g. Wilson type)
fermion discretization.

\section{Acknowledgments}
Computations were performed on the BlueGene
at FZ J\"ulich and on clusters at Wuppertal
and Budapest equipped with graphics cards. 
This work is supported in part by Hungarian OTKA
grant AT04965, DFG grants SFB-TR 55, FO 502/1-2, EU grant   
(FP7/2007-2013)/ERC n$^o$208740. 

\bibliographystyle{JHEP}
\bibliography{newtc}

\providecommand{\href}[2]{#2}\begingroup\raggedright\begin{thebibliography}{10}

\bibitem{Aoki:2006we}
Y.~Aoki, G.~Endrodi, Z.~Fodor, S.~D. Katz, and K.~K. Szabo, {\it {The order of
  the quantum chromodynamics transition predicted by the standard model of
  particle physics}},  {\em Nature} {\bf 443} (2006) 675--678,
  [\href{http://xxx.lanl.gov/abs/hep-lat/0611014}{{\tt hep-lat/0611014}}].

\bibitem{Cheng:2006qk}
M.~Cheng {\em et.~al.}, {\it {The transition temperature in QCD}},  {\em Phys.
  Rev.} {\bf D74} (2006) 054507,
  [\href{http://xxx.lanl.gov/abs/hep-lat/0608013}{{\tt hep-lat/0608013}}].

\bibitem{Detar:2007as}
{\bf HotQCD} Collaboration, C.~Detar and R.~Gupta, {\it {Toward a precise
  determination of Tc with 2+1 flavors of quarks}},  {\em PoS} {\bf LAT2007}
  (2007) 179, [\href{http://xxx.lanl.gov/abs/0710.1655}{{\tt
  arXiv:0710.1655}}].

\bibitem{Karsch:2007dt}
F.~Karsch, {\it {Recent lattice results on finite temerature and density QCD,
  part II}},  {\em PoS} {\bf LAT2007} (2007) 015,
  [\href{http://xxx.lanl.gov/abs/0711.0661}{{\tt arXiv:0711.0661}}].

\bibitem{Karsch:2008fe}
{\bf RBC} Collaboration, F.~Karsch, {\it {Equation of state and more from
  lattice regularized QCD}},  \href{http://xxx.lanl.gov/abs/0804.4148}{{\tt
  arXiv:0804.4148}}.

\bibitem{Aoki:2006br}
Y.~Aoki, Z.~Fodor, S.~D. Katz, and K.~K. Szabo, {\it {The QCD transition
  temperature: Results with physical masses in the continuum limit}},  {\em
  Phys. Lett.} {\bf B643} (2006) 46--54,
  [\href{http://xxx.lanl.gov/abs/hep-lat/0609068}{{\tt hep-lat/0609068}}].

\bibitem{Bazavov:2007zz}
A.~Bazavov and B.~A. Berg, {\it {Deconfining Phase Transition on Lattices with
  Boundaries at Low Temperature}},  {\em Phys. Rev.} {\bf D76} (2007) 014502,
  [\href{http://xxx.lanl.gov/abs/hep-lat/0701007}{{\tt hep-lat/0701007}}].

\bibitem{Takaishi:2005tz}
T.~Takaishi and P.~de~Forcrand, {\it {Testing and tuning new symplectic
  integrators for hybrid Monte Carlo algorithm in lattice QCD}},  {\em Phys.
  Rev.} {\bf E73} (2006) 036706,
  [\href{http://xxx.lanl.gov/abs/hep-lat/0505020}{{\tt hep-lat/0505020}}].

\bibitem{Clark:2006fx}
M.~A. Clark and A.~D. Kennedy, {\it {Accelerating dynamical fermion
  computations using the rational hybrid Monte Carlo (RHMC) algorithm with
  multiple pseudofermion fields}},  {\em Phys. Rev. Lett.} {\bf 98} (2007)
  051601, [\href{http://xxx.lanl.gov/abs/hep-lat/0608015}{{\tt
  hep-lat/0608015}}].

\bibitem{Egri:2006zm}
G.~I. Egri {\em et.~al.}, {\it {Lattice QCD as a video game}},  {\em Comput.
  Phys. Commun.} {\bf 177} (2007) 631--639,
  [\href{http://xxx.lanl.gov/abs/hep-lat/0611022}{{\tt hep-lat/0611022}}].

\bibitem{Colangelo:2005gd}
G.~Colangelo, S.~Durr, and C.~Haefeli, {\it {Finite volume effects for meson
  masses and decay constants}},  {\em Nucl. Phys.} {\bf B721} (2005) 136--174,
  [\href{http://xxx.lanl.gov/abs/hep-lat/0503014}{{\tt hep-lat/0503014}}].

\bibitem{Ishizuka:1993mt}
N.~Ishizuka, M.~Fukugita, H.~Mino, M.~Okawa, and A.~Ukawa, {\it {Operator
  dependence of hadron masses for Kogut-Susskind quarks on the lattice}},  {\em
  Nucl. Phys.} {\bf B411} (1994) 875--902.

\bibitem{Bernard:2001av}
C.~W. Bernard {\em et.~al.}, {\it {The QCD spectrum with three quark flavors}},
   {\em Phys. Rev.} {\bf D64} (2001) 054506,
  [\href{http://xxx.lanl.gov/abs/hep-lat/0104002}{{\tt hep-lat/0104002}}].

\bibitem{Amsler:2008zz}
{\bf Particle Data Group} Collaboration, C.~Amsler {\em et.~al.}, {\it {Review
  of particle physics}},  {\em Phys. Lett.} {\bf B667} (2008) 1.

\bibitem{Eidelman:2004wy}
{\bf Particle Data Group} Collaboration, S.~Eidelman {\em et.~al.}, {\it
  {Review of particle physics}},  {\em Phys. Lett.} {\bf B592} (2004) 1.

\bibitem{Aoki:2008sm}
{\bf PACS-CS} Collaboration, S.~Aoki {\em et.~al.}, {\it {2+1 Flavor Lattice
  QCD toward the Physical Point}},
  \href{http://xxx.lanl.gov/abs/0807.1661}{{\tt arXiv:0807.1661}}.

\bibitem{Aoki:2099}
Y.~Aoki, S.~Durr, Z.~Fodor, S.~D. Katz, S.~Krieg, and K.~K. Szabo, {\it {Quark
  masses and decay constants from staggered simulations at the physical
  point}},  {\em In preparation}.

\bibitem{Gray:2005ur}
A.~Gray {\em et.~al.}, {\it {The Upsilon spectrum and m(b) from full lattice
  QCD}},  {\em Phys. Rev.} {\bf D72} (2005) 094507,
  [\href{http://xxx.lanl.gov/abs/hep-lat/0507013}{{\tt hep-lat/0507013}}].

\bibitem{Sommer:1993ce}
R.~Sommer, {\it {A New way to set the energy scale in lattice gauge theories
  and its applications to the static force and alpha-s in SU(2) Yang-Mills
  theory}},  {\em Nucl. Phys.} {\bf B411} (1994) 839--854,
  [\href{http://xxx.lanl.gov/abs/hep-lat/9310022}{{\tt hep-lat/9310022}}].

\bibitem{Albanese:1987ds}
{\bf APE} Collaboration, M.~Albanese {\em et.~al.}, {\it {Glueball Masses and
  String Tension in Lattice QCD}},  {\em Phys. Lett.} {\bf B192} (1987) 163.

\bibitem{Hasenfratz:2001hp}
A.~Hasenfratz and F.~Knechtli, {\it {Flavor symmetry and the static potential
  with hypercubic blocking}},  {\em Phys. Rev.} {\bf D64} (2001) 034504,
  [\href{http://xxx.lanl.gov/abs/hep-lat/0103029}{{\tt hep-lat/0103029}}].

\bibitem{Gockeler:2005rv}
M.~Gockeler {\em et.~al.}, {\it {A determination of the Lambda parameter from
  full lattice QCD}},  {\em Phys. Rev.} {\bf D73} (2006) 014513,
  [\href{http://xxx.lanl.gov/abs/hep-ph/0502212}{{\tt hep-ph/0502212}}].

\bibitem{Borsanyi:2099}
S.~Borsanyi {\em et.~al.}, {\it {Lines of constant physics for arbitrary small
  lattice spacings}},  {\em In preparation}.

\bibitem{Soltz:2008}
R.~Soltz, {\it {Recent Results from HotQCD and Experimental Benchmarks for
  Hydrodynamnics}},  {\em RIKEN BNL Research Center Workshop Apr 2008}
  http://quark.phy.bnl.gov/~petreczk/ideal\_hydro/monday/bnlhydro08\_soltz.ppt.

\bibitem{Cheng:2007jq}
M.~Cheng {\em et.~al.}, {\it {The QCD Equation of State with almost Physical
  Quark Masses}},  {\em Phys. Rev.} {\bf D77} (2008) 014511,
  [\href{http://xxx.lanl.gov/abs/0710.0354}{{\tt arXiv:0710.0354}}].

\bibitem{Karsch:2007dp}
F.~Karsch, {\it {Recent lattice results on finite temperature and density QCD,
  part I}},  {\em PoS} {\bf CPOD07} (2007) 026,
  [\href{http://xxx.lanl.gov/abs/0711.0656}{{\tt arXiv:0711.0656}}].

\end{thebibliography}\endgroup
\end{document}